\documentclass[12pt]{article}	
\usepackage{mathrsfs, amsfonts, subeqn}

\font\hb=bbmsl10 scaled 1200

\def\Eqn#1{Eq.\ (\ref{#1})}
\def\Eqs#1#2{Eqs.\ (\ref{#1}) and (\ref{#2})}
\def\3Eqs#1#2#3{Eqs.\ (\ref{#1}), (\ref{#2}) and (\ref{#3})}
\def\sec#1{\S\,\ref{#1}}
\def\vec#1{\mathchoice
{\mbox{\boldmath $#1$}}
{\mbox{\boldmath $#1$}}
{\mbox{\boldmath $\scriptstyle #1$}}
{\mbox{\boldmath $\scriptscriptstyle #1$}}}
\def\(({(\!(}
\def\)){)\!)}
\newcommand{\tr}{\mathop{\rm Tr}}
\def\umat{\mbox{\hb 1}}

\makeatletter
\renewcommand{\theequation}{\thesection.\arabic{equation}}
\@addtoreset{equation}{section}
\makeatother

\newtheorem{theorem}{Theorem}

\title{\bf Representation-independent\\ manipulations with\\ Dirac
  matrices and spinors} 

\author{\bf Palash B. Pal\\ 
\normalsize Saha Institute of Nuclear Physics\\ 
\normalsize 1/AF Bidhan-Nagar, Calcutta 700064, INDIA}

\date{}

\begin{document}

\maketitle

\begin{abstract}
Dirac matrices, also known as gamma matrices, are defined only up to a
similarity transformation.  Usually, some explicit representation of
these matrices is assumed in order to deal with them.  In this
article, we show how it is possible to proceed without any explicit
form of these matrices.  Various important identities involving Dirac
matrices and spinors have been derived without assuming any
representation at any stage.

\centerline{\tt PACS numbers: 03.65.Pm, 31.30.jx}

\end{abstract}


\section{Introduction}\label{s:in}
In order to obtain a relativistically covariant equation for the
quantum mechanical wave function, Dirac introduced a Hamiltonian that
is linear in the momentum operator.  In modern notation, it can be
written as
\begin{eqnarray}
H = \gamma^0 \Big( \vec \gamma \cdot \vec p_{\rm op} +  m \Big) \,,
\label{H}
\end{eqnarray}
where $m$ is the mass of the particle and $\vec p_{\rm op}$ the
momentum operator.  We will throughout use natural units with
$c=\hbar=1$ so that $\gamma^0$ and $\vec\gamma$ are dimensionless.
Because of their anticommutation properties that we mention in
\sec{s:bp}, they have to be matrices.  The four matrices are written
together as
\begin{eqnarray}
\gamma^\mu \equiv \{ \gamma^0 , \gamma^i \} \,,
\end{eqnarray}
where we have put a Lorentz index in the left hand side.  We will also
define the corresponding matrices with lower indices in the usual way:
\begin{eqnarray}
\gamma_\mu = g_{\mu\nu} \gamma^\nu \,,
\end{eqnarray}
where $g_{\mu\nu}$ is the metric tensor, for which our convention has
been stated in \Eqn{g}.  Of course, the Lorentz indices on the gamma
matrices do not imply that the matrices transform as vectors.  They
are, in fact, constant matrices which are frame-independent.  The
Lorentz index in $\gamma^\mu$ only indicates that the four quantities
obtained by sandwiching these matrices between fermionic fields
transform as components of a vector.

Some properties of the Dirac matrices follow directly from their
definition in \Eqn{H}, as shown in \sec{s:bp}.  However, these
properties do not specify the elements of the matrices uniquely.  They
only define the matrices up to a similarity transformation.
Since spinors are plane-wave solutions of the equation 
\begin{eqnarray}
i {\partial\psi \over \partial t} = H\psi \,,
\label{SchEq}
\end{eqnarray}
and $H$ contains the Dirac matrices which are not uniquely defined,
the solutions also share this non-uniqueness.

In physics, whenever there is an arbitrariness in the definition of
some quantity, it is considered best to deal with combinations of
those quantities which do not suffer from the arbitrariness.  For
example, components of a vector depend on the choice of the axes of
co-ordinates.  Physically meaningful relations can either involve
things like scalar products of vectors which do not depend on the
choice of axes, or are in the form of equality of two quantities (say,
two vectors) both of which transform the same way under a rotation of
the axes, so that their equality is not affected.  Needless to say, it
is best if we can follow the same principles while dealing with Dirac
matrices and spinors.  However, in most texts dealing with them, this
approach is not taken \cite{texts}.  Most frequently, one chooses an
explicit representation of the Dirac matrices and spinors, and works
with it.

Apart from the fact that an explicit representation is aesthetically
less satisfying, it must also be said that dealing with them can also
lead to pitfalls.  One might use some relation which holds in some
specific representation but not in general, and obtain a wrong
conclusion.

In this article, we show how, without using any explicit
representation of the Dirac matrices or spinors, one can obtain useful
relations involving them.  The article is organized as follows.  In
\sec{s:bp}, we define the basic properties of Dirac matrices and
spinors and mention the extent of arbitrariness in the definitions.
In \sec{s:as}, we recall some well-known associated matrices which are
useful in dealing with Dirac matrices.  In \sec{s:im}, we derive some
identities involving the Dirac matrices and associated matrices in a
completely representation-independent way.  In \sec{s:sp}, we show how
spinor solutions can be defined in a representation-independent
fashion and identify their combinations on which normalization
conditions can be imposed.  We derive some important relations
involving spinors in \sec{s:rl}, and involving spinor biliears in
\sec{s:bl}.  Concluding remarks appear in \sec{s:cr}.

\section{Basic properties of Dirac matrices and spinors}\label{s:bp}
Some properties of the Dirac matrices are immediately derived from
\Eqn{H}.  First, the relativistic Hamiltonian of a free particle is
given by
\begin{eqnarray}
H^2 = \vec p^2 + m^2 \,,
\end{eqnarray}
and \Eqn{H}, when squared, must yield this relation.  Assuming
$\gamma_0$ and $\vec\gamma$ commute with the momentum operator, this
gives a set of relations which can be summarized in the form
\begin{eqnarray}
\Big[ \gamma_\mu , \gamma_\nu \Big]_+ &=& 2 g_{\mu\nu} \umat \,,
\label{acomm}
\end{eqnarray}
where $g_{\mu\nu}$ is the metric defined in \Eqn{g}, and $\umat$ is
the unit matrix which will not be always explicitly written in the
subsequent formulas.  This relation requires that the Dirac matrices
are at least $4\times4$ matrices, and we take them to be $4\times4$.

Hermiticity of the Hamiltonian of \Eqn{H} gives some further
conditions on the Dirac matrices, namely that $\gamma_0$ must be
hermitian, and so should be the combinations $\gamma_0\gamma_i$.  Both
these relations can be summarized by writing
\begin{eqnarray}
\gamma_\mu^\dagger = \gamma_0 \gamma_\mu \gamma_0 
\label{gamdag}
\end{eqnarray}
in view of the anticommutation relations given in \Eqn{acomm}.

\Eqs{acomm}{gamdag} are the basic properties which define the Dirac
matrices.  With these defining relations, the arbitrariness can be
easily seen through the following theorems.
\begin{theorem}\label{th:U}
For any choice of the matrices $\gamma_\mu$ satisfying
\Eqs{acomm}{gamdag}, if we take another set defined by
\begin{eqnarray}
\tilde\gamma_\mu = U \gamma_\mu U^\dagger
\label{gamtil}
\end{eqnarray}
for some unitary matrix $U$, then these new matrices satisfy the same
anticommutation and hermiticity properties as the matrices
$\gamma_\mu$. 
\end{theorem}
The proof of this theorem is straight forward and trivial.  The
converse is also true:
\begin{theorem}\label{th:Uconv}
If two sets of matrices $\gamma_\mu$ and
$\tilde\gamma_\mu$ both satisfy \Eqs{acomm}{gamdag}, they are related
through \Eqn{gamtil} for some unitary matrix $U$.
\end{theorem}
The proof is non-trivial \cite{schweber,J&R} and we will not give it
here.  The two theorems show that the Dirac matrices are defined only
up to a similarity transformation with a unitary matrix.

To obtain the defining equation for the spinors, we multiply both
sides of \Eqn{SchEq} by $\gamma_0$ and put $\vec p_{\rm
op}=-i\vec\nabla$ into the Hamiltonian of \Eqn{H}.  This gives the
Dirac equation:
\begin{eqnarray}
i \gamma^\mu \partial_\mu \psi - m\psi = 0 \,.
\label{DiracEq}
\end{eqnarray}
There are two types of plane-wave solutions:
\begin{eqnarray}
\psi \sim 
\cases {u\((\vec p\)) e^{-ip\cdot x} \,, \cr 
v\((\vec p\)) e^{+ip\cdot x} \,. }
\label{planewave}
\end{eqnarray}
Here and later, we indicate functional dependence in double
parentheses so that it does not get confused with multiplicative
factors in parentheses.  The objects $u\((\vec p\))$ and $v\((\vec
p\))$ are 4-component column vectors, and will be called ``spinors''.
The 4-vector $p^\mu$ is given by
\begin{eqnarray}
p^\mu \equiv \{ E_{\vec p}, \vec p \} \,,
\end{eqnarray}
where $E_{\vec p}$ is the positive energy eigenvalue:
\begin{eqnarray}
E_{\vec p} = + \sqrt {\vec p^2 + m^2} \,.
\label{Ep}
\end{eqnarray}

Putting \Eqn{planewave} into \Eqn{DiracEq}, we obtain the equations that
define the $u$ and $v$-spinors:
\begin{subequations}
\label{uveqn}
\begin{eqnarray}
( \gamma_\mu p^\mu - m ) u\((\vec p\)) &=& 0 \,, 
\label{ueqn} \\
( \gamma_\mu p^\mu + m ) v\((\vec p\)) &=& 0 \,.
\label{veqn}
\end{eqnarray}
\end{subequations}
Obviously, if we change $\gamma^\mu$ to $\tilde\gamma_\mu$ through the
prescription given in \Eqn{gamtil} and also change the spinors to
\begin{eqnarray}
\tilde u\((\vec p\)) = U u\((\vec p\)) \,, \qquad 
\tilde v\((\vec p\)) = U v\((\vec p\)) \,, 
\label{Uuv}
\end{eqnarray}
\Eqn{uveqn} is satisfied by the new matrices and the new
spinors.  \Eqn{Uuv} shows that the spinors themselves are
representation-dependent.

\section{Some associated matrices}\label{s:as}
In order to proceed, we recall the definitions of some matrices
associated with the Dirac matrices.  These definitions can be obtained
in any textbook dealing with Dirac particles or fields, but are
compiled here for the sake of completeness.

The sigma-matrices are defined as
\begin{eqnarray}
\sigma_{\mu\nu} = {i \over 2} \Big[ \gamma_\mu , \gamma_\nu \Big] \,.
\label{sigdef}
\end{eqnarray}
The matrices $\frac12\sigma_{\mu\nu}$ constitute a representation of
the Lorentz group.  The subgroup of rotation group has the generators
$\frac12\sigma_{ij}$, with both spatial indices.  We define the spin
matrices:
\begin{eqnarray}
\Sigma^i = \frac12 \varepsilon^{ijk} \sigma_{jk} \,,
\label{Sigma}
\end{eqnarray}
so that $\frac12\Sigma^i$ represent the spin components.  From
\Eqn{gamdag}, it is easy to check that the matrices $\sigma_{0i}$ are
anti-hermitian, whereas the matrices $\sigma_{ij}$, and therefore the
matrices $\Sigma^i$, are hermitian.

The next important matrix is defined from the observation that the
matrices $-\gamma_\mu^\top$ satisfy the same anticommutation and
hermiticity properties as $\gamma_\mu$.  By Theorem~\ref{th:Uconv},
there must then exist a unitary matrix $C$, 
\begin{eqnarray}
C^\dagger = C^{-1} \,,
\end{eqnarray}
such that
\begin{eqnarray}
C^{-1} \gamma_\mu C = - \gamma_\mu^\top \,.
\label{C}
\end{eqnarray}
Note that the definitions in \Eqs{sigdef}{C} imply the relation
\begin{eqnarray}
C^{-1} \sigma_{\mu\nu} C = - \sigma_{\mu\nu}^\top \,.
\label{CsigCinv}
\end{eqnarray}

Another important matrix is $\gamma_5$, defined as
\begin{eqnarray}
\gamma_5 = i \gamma^0 \gamma^1 \gamma^2 \gamma^3 \,,
\label{g5def}
\end{eqnarray}
or equivalently as
\begin{eqnarray}
\gamma_5 = {i \over 4!} \; \varepsilon_{\mu\nu\lambda\rho} \gamma^\mu
\gamma^\nu \gamma^\lambda \gamma^\rho \,,
\label{gam5alt}
\end{eqnarray}
where $\varepsilon_{\mu\nu\lambda\rho}$ stands for the completely
antisymmetric rank-4 tensor, with
\begin{eqnarray}
\varepsilon_{0123} = 1 \,.
\label{LCconv}
\end{eqnarray}

From \Eqn{acomm}, it is easily seen that 
\begin{eqnarray}
\Big(\gamma_5\Big)^2 = \umat \,.
\label{g5sq}
\end{eqnarray}
It is also easy to see that $\gamma_5$ anticommutes with all
$\gamma_\mu$'s and commutes with all $\sigma_{\mu\nu}$'s:
\begin{eqnarray}
\Big[ \gamma_\mu , \gamma_5 \Big]_+ &=& 0 \,, 
\label{g5gmu}\\*
\Big[ \sigma_{\mu\nu} , \gamma_5 \Big] &=& 0 \,.
\label{g5sigma}
\end{eqnarray}
Also, \Eqs{g5def}{C} imply the relation
\begin{eqnarray}
C^{-1} \gamma_5 C = \gamma_5^\top \,.
\label{Cinvg5C}
\end{eqnarray}

There is another property of $\gamma_5$ which is of interest.
Consider the trace of this matrix.  Using the anticommutation
properties of the Dirac matrices, we can write
\begin{eqnarray}
\tr \Big( \gamma_5 \Big) &=& i \tr (\gamma^0 \gamma^1 \gamma^2 \gamma^3)
= - i \tr (\gamma^0 \gamma^1 \gamma^3 \gamma^2) \nonumber\\ 
&=& + i \tr (\gamma^0 \gamma^3 \gamma^1 \gamma^2)
= - i \tr (\gamma^3 \gamma^0 \gamma^1 \gamma^2) \,.
\end{eqnarray}
Now, since $\tr(M_1M_2) = \tr(M_2M_1)$ for any two matrices $M_1$ and
$M_2$, we can take $M_1=\gamma^3$ and $M_2=\gamma^0 \gamma^1
\gamma^2$, use this cyclic property, and get
\begin{eqnarray}
\tr \Big( \gamma_5 \Big) = - i \tr (\gamma^0 \gamma^1 \gamma^2
\gamma^3) = - \tr \Big( \gamma_5 \Big) \,,
\end{eqnarray}
which means that
\begin{eqnarray}
\tr \Big( \gamma_5 \Big) = 0 \,.
\label{trgam5}
\end{eqnarray}

Using the argument that leads to the existence of the matrix $C$, we
can define some other associated matrices \cite{J&R}.  For example,
given any representation of the matrices $\gamma_\mu$, the matrices
$\gamma_\mu^\dagger$ also satisfy the same anticommutation and
hermiticity properties as the matrices $\gamma_\mu$, and therefore
there must be a unitary matrix $A$ that satisfies the relation
\begin{subequations}
\label{ABCD}
\begin{eqnarray}
A^{-1} \gamma_\mu A = \gamma_\mu^\dagger \,.
\label{A}
\end{eqnarray}
Similarly, there exist unitary matrices $A'$, $B$, $B'$ and $C'$ which
satisfy the relations
\begin{eqnarray}
A'^{-1} \gamma_\mu A' = - \gamma_\mu^\dagger \,, 
\label{A'}\\ 
C'^{-1} \gamma_\mu C' = \gamma_\mu^\top \,, 
\label{C'}\\ 
B^{-1} \gamma_\mu B = \gamma_\mu^* \,, 
\label{B}\\ 
B'^{-1} \gamma_\mu B' = -\gamma_\mu^* 
\label{B'}\,. 
\end{eqnarray}
\end{subequations}
Fortunately, there is no need to discuss these matrices separately.
\Eqn{gamdag}, used in \Eqn{A}, gives the relation
\begin{eqnarray}
\Big[ \gamma_\mu, A \gamma_0 \Big] = 0 \,.
\end{eqnarray}
Thus $A\gamma_0$ must be a multiple of the unit matrix, i.e.,
$A=\alpha \gamma_0$ for some number $\alpha$.  Since $A$ will have to
be unitary, $\alpha$ can only be a phase.  Thus we conclude that 
\begin{eqnarray}
A = \gamma_0
\end{eqnarray}
apart from a possible overall phase.  Similarly, we can see that
$A'\gamma_0$ must anticommute with any $\gamma_\mu$, and so we must
have 
\begin{eqnarray}
A' = \gamma_5 \gamma_0
\end{eqnarray}
up to a phase factor.  By a similar argument, we find
\begin{eqnarray}
C' = \gamma_5 C 
\end{eqnarray}
up to an overall phase.  As for $B$ and $B'$, we note that
\begin{eqnarray}
\gamma_\mu^* = (\gamma_\mu^\dagger)^\top = \Big( \gamma_0 \gamma_\mu
\gamma_0 \Big)^\top = - C^{-1} \gamma_0 \gamma_\mu
\gamma_0 C = - (\gamma_0 C)^{-1} \gamma_\mu (\gamma_0 C) \,.
\end{eqnarray}
Thus, up to overall phase factors, one obtains
\begin{subequations}
\begin{eqnarray}
B' &=& \gamma_0 C \,, \\ 
B &=& \gamma_5 \gamma_0 C \,.
\end{eqnarray}
\end{subequations}
The following properties follow trivially, using the explicit form for
the matrix $B'$:
\begin{subequations}
\begin{eqnarray}
\sigma_{\mu\nu}^* &=& (\gamma_0 C)^{-1} \sigma_{\mu\nu} (\gamma_0 C)
\,, \\ 
\gamma_5^* &=& - (\gamma_0 C)^{-1} \gamma_5 (\gamma_0 C)
\,, \\
(\gamma_\mu \gamma_5)^* &=& (\gamma_0 C)^{-1} \gamma_\mu \gamma_5
(\gamma_0 C) \,.
\end{eqnarray}
\end{subequations}
%

\section{Identities involving Dirac matrices}\label{s:im}
\subsection{Trace identities}
Previously, we have shown that the matrix $\gamma_5$ is traceless.
We can also try to find the trace of any of the Dirac matrices
$\gamma_\mu$.  Using \Eqn{g5sq}, we can write
\begin{eqnarray}
\tr \Big( \gamma_\mu \Big) = \tr \Big( \gamma_\mu \gamma_5 \gamma_5
\Big) \,.
\end{eqnarray}
Then, using the cyclic property of traces and \Eqn{g5gmu}, we obtain
\begin{eqnarray}
\tr \Big( \gamma_\mu \Big) = \tr \Big( \gamma_5 \gamma_\mu \gamma_5
\Big) = - \tr \Big( \gamma_\mu \gamma_5 \gamma_5 \Big) \,.
\end{eqnarray}
Comparing the two equations, we obtain
\begin{eqnarray}
\tr \Big( \gamma_\mu \Big) = 0 \,.
\label{0trace}
\end{eqnarray}
The same technique can be employed to prove that the trace of the
product of any odd number of Dirac matrices is zero.  For the product
of even number of Dirac matrices, we can use the result
\begin{eqnarray}
\tr \Big( \gamma_\mu \gamma_\nu \Big) &=& 4g_{\mu\nu} \,, \\
\tr \Big( \gamma_\mu \gamma_\nu \gamma_\lambda \gamma_\rho \Big) &=&
4 \Big( g_{\mu\nu} g_{\lambda\rho} - g_{\mu\lambda} g_{\nu\rho} +
g_{\mu\rho} g_{\nu\lambda} \Big)  \,, 
\end{eqnarray}
and so on.  We do not give the details of the proofs because they are
usually proved in a representation-independent manner in textbooks.

\subsection{Contraction identities}
First, there are the contraction formulas involving only the Dirac
matrices, e.g.,
\begin{subequations}
\label{contr}
\begin{eqnarray}
\gamma^\mu \gamma_\mu &=& 4 \,,
\label{contr2}
\\ 
\gamma^\mu \gamma_\nu \gamma_\mu &=& - 2 \gamma_\nu \,,
\label{contr3}
\\ 
\gamma^\mu \gamma_\nu \gamma_\lambda \gamma_\mu &=& 4 g_{\nu\lambda} \,,
\label{contr4}
\\ 
\gamma^\mu \gamma_\nu \gamma_\lambda \gamma_\rho \gamma_\mu &=& - 2
\gamma_\rho \gamma_\lambda \gamma_\nu \,, 
\label{contr5}
\end{eqnarray}
\end{subequations}
and so on for longer strings of Dirac matrices, which can be proved
easily by using the anticommutation relation of \Eqn{acomm}.  There
are also similar formulas involving contractions of the sigma
matrices, like
\begin{subequations}
\label{sigmas}
\begin{eqnarray}
\sigma^{\mu\nu} \sigma_{\mu\nu} &=& 12 \,,
\label{sigsig} \\ 
\sigma^{\mu\nu} \sigma^{\lambda\rho} \sigma_{\mu\nu} &=& - 4
\sigma^{\lambda\rho} \,, 
\label{sigsigsig}  
\end{eqnarray}
\end{subequations}
and some other involving both gamma matrices and sigma matrices:
\begin{subequations}
\label{sig&gam}
\begin{eqnarray}
\sigma^{\mu\nu} \gamma^\lambda \sigma_{\mu\nu} &=& 0 \,,
\label{siggamsig} \\ 
\gamma^\lambda \sigma^{\mu\nu} \gamma_\lambda &=& 0 \,.
\label{gamsiggam}
\end{eqnarray}
\end{subequations}
All of these can be easily proved by using the definition of the sigma
matrices and the contraction formulas for the gamma matrices.

\subsection{Identities from linear independence}
There are many other identities involving the Dirac matrices which
are derived from the fact that the 16 matrices
\begin{eqnarray}
\umat, \gamma_\mu, \sigma_{\mu\nu} \mbox{(for $\mu<\nu$)},
\gamma_\mu\gamma_5, \gamma_5
\label{basis16}
\end{eqnarray}
constitute a complete set of $4\times4$ matrices.  In other words, any
$4\times4$ matrix $M$ can be expressed as a linear superposition of
these 16 matrices:  
\begin{eqnarray}
M = a \umat + b^\mu \gamma_\mu + c^{\mu\nu} \sigma_{\mu\nu} 
+ d^\mu \gamma_\mu\gamma_5 + e \gamma_5 \,.
\label{M}
\end{eqnarray}

In particular, any product of any number of these basis matrices can
also be written in the form proposed in \Eqn{M}, with suitable choices
of the co-efficients $a$, $b^\mu$, $c^{\mu\nu}$, $d^\mu$ and $e$.  One
example of this kind of relation is the identity
\begin{eqnarray}
\gamma_\mu \gamma_\nu = g_{\mu\nu} \umat - i \sigma_{\mu\nu} \,,
\label{gamgam}
\end{eqnarray}
which follows trivially from \Eqs{acomm}{sigdef}.
\Eqs{sigmas}{sig&gam} are also examples of this general
theme.  To see more examples of this kind, let us consider the
combination $\varepsilon^{\mu\nu\lambda\rho} \sigma_{\mu\nu}
\gamma_5$.  We can use the definition of $\gamma_5$ from
\Eqn{gam5alt}, and use the product of two Levi-Civita symbols given in
\Eqn{epseps}.  This gives
\begin{eqnarray}
\varepsilon^{\mu\nu\lambda\rho} \sigma_{\mu\nu} \gamma_5 
= - {i \over 4!} \; \sigma_{\mu\nu} \Big( \gamma^\mu \gamma^\nu
\gamma^\lambda \gamma^\rho + (-1)^P \mbox{(permutations)} \Big) \,,
\end{eqnarray}
where the factor $(-1)^P$ is $+1$ if the permutation is even, and $-1$
if the permutation is odd.  There are 24 possible permutations.  Each
of them can be simplified by using one or other of the contraction
formulas given above, and the result is
\begin{eqnarray}
\varepsilon^{\mu\nu\lambda\rho} \sigma_{\mu\nu} \gamma_5 = 2i
\sigma^{\lambda\rho} \,,
\label{epssig5}
\end{eqnarray}
or equivalently
\begin{eqnarray}
\sigma^{\lambda\rho} \gamma_5 = - {i \over 2}
\varepsilon^{\mu\nu\lambda\rho} \sigma_{\mu\nu} \,.
\label{sig5}
\end{eqnarray}

A very useful identity can be derived by starting with the combination
$\varepsilon_{\mu\nu\lambda\rho} \gamma^\rho\gamma_5$, and using
\Eqs{gam5alt}{epseps}, as was done for deducing \Eqn{epssig5}.  The
final result can be expressed in the form
\begin{eqnarray}
\gamma_\mu \gamma_\nu \gamma_\lambda = g_{\mu\nu} \gamma_\lambda +
g_{\nu\lambda} \gamma_\mu - g_{\lambda\mu} \gamma_\nu - i
\varepsilon_{\mu\nu\lambda\rho} \gamma^\rho\gamma_5 \,.
\label{ggg}
\end{eqnarray}
With this identity, any string of three or more gamma matrices can be
reduced to strings of smaller number of gamma matrices.

An important identity can be derived by multiplying \Eqn{gamgam} by
$\gamma_5$, and using \Eqn{sig5}.  This gives
\begin{eqnarray}
\gamma_\mu \gamma_\nu \gamma_5 = g_{\mu\nu} \gamma_5 - \frac12 
\varepsilon_{\mu\nu\lambda\rho} \sigma^{\lambda\rho} \,.
\end{eqnarray}
In particular, if the index $\mu$ is taken to be in the time
direction and the index $\nu$ to be a spatial index,  we obtain
\begin{eqnarray}
\gamma_0 \gamma_i \gamma_5 = - \frac12 
\varepsilon_{0ijk} \sigma^{jk} \,.
\end{eqnarray}
Taking the convention for the completely antisymmetric 3-dimensional
tensor in such a way that $\varepsilon_{0ijk} = \varepsilon_{ijk}$, we
can rewrite this equation by comparing the right hand side with the
definition of the spin matrices in \Eqn{Sigma}:
\begin{eqnarray}
\Sigma_i = - \gamma_0 \gamma_i \gamma_5 \,. 
\label{0i5}
\end{eqnarray}
With this form, it is easy to show that
\begin{eqnarray}
\Big[ \Sigma_i , \Sigma_j \Big]_+ = 2 \delta_{ij} \,,
\end{eqnarray}
by using anticommutation properties of the gamma matrices.

\subsection{Antisymmetry of $C$}\label{ss:Ctrans}
Taking the transpose of \Eqn{C} that defines  the matrix $C$, we
obtain 
\begin{eqnarray}
\gamma_\mu = - C^\top \gamma_\mu^\top (C^{-1})^\top =  C^\top
C^{-1} \gamma_\mu C (C^{-1})^\top =  C^\top
C^{-1} \gamma_\mu (C^\top  C^{-1})^{-1} \,,
\end{eqnarray}
which can be rewritten in the form
\begin{eqnarray}
\Big[ \gamma_\mu ,  C^\top  C^{-1} \Big] = 0 \,.
\end{eqnarray}
Because the matrix $C^\top  C^{-1}$ commutes with each Dirac matrix,
it must be a multiple of the unit matrix.  So we write
\begin{eqnarray}
C^\top = \lambda C 
\label{lambdaC}
\end{eqnarray}
for some number $\lambda$.  Taking transpose of both sides of this
equation, we obtain
\begin{eqnarray}
C = \lambda C^\top \,,
\end{eqnarray}
and therefore $\lambda^2=1$.

We now go back to \Eqn{C} and rewrite it in the form
\begin{eqnarray}
\gamma_\mu C = - C \gamma_\mu^\top = -\lambda (\gamma_\mu C)^\top \,.
\end{eqnarray}
This means that the matrices $\gamma_\mu C$ are all antisymmetric if
$\lambda=+1$, and symmetric if $\lambda=-1$.  Using the definition of
the sigma-matrices from \Eqn{sigdef}, it is easy to show that the
matrices $\sigma_{\mu\nu} C$ have the same properties as well:
\begin{eqnarray}
\sigma_{\mu\nu} C = -\lambda ( \sigma_{\mu\nu} C )^\top \,.
\end{eqnarray}
In addition, note that the four matrices $\gamma_\mu C$ and the six
matrices $\sigma_{\mu\nu} C$ are linearly independent, because the ten
matrices $\gamma_\mu$ and $\sigma_{\mu\nu}$ are.  There cannot be ten
linearly independent antisymmetric $4\times4$ matrices.  Thus, the
matrices $\gamma_\mu C$ and $\sigma_{\mu\nu} C$ must be all symmetric,
implying 
\begin{eqnarray}
\lambda = -1 \,,
\end{eqnarray}
i.e.,
\begin{eqnarray}
C^\top = -C \,.
\label{CT=-C}
\end{eqnarray}
The matrix $C$ is therefore antisymmetric~\cite{Ctrans}.

Once the choice of $\lambda$ has been determined, it is easy to see,
using \Eqn{Cinvg5C}, that the matrix $\gamma_5C$ and the four matrices
$\gamma_\mu\gamma_5C$ are antisymmetric.  Thus, any symmetric
$4\times4$ matrix can be written as a linear superposition of the ten
matrices $\gamma_\mu C$ and $\sigma_{\mu\nu} C$, whereas any
antisymmetric $4\times4$ matrix can be written as a linear
superposition of the six matrices $C$, $\gamma_5 C$ and $\gamma_\mu
\gamma_5 C$.  In this sense, this collection of 16 matrices is a
better basis, compared to that given in \Eqn{basis16}, for writing an
arbitrary $4\times4$ matrix.

\section{Spinors}\label{s:sp}
\subsection{Eigenvectors of $\gamma_0$}
Consider the matrix $\gamma_0$.  It is a $4\times4$ matrix, so it has
four eigenvalues and eigenvectors.  It is hermitian, so the
eigenvalues are real.  In fact, from \Eqn{acomm} we know that its
square is the unit matrix, so that its eigenvalues can only be $\pm1$.
Since $\gamma_0$ is traceless, as we have proved in \Eqn{0trace},
there must be two eigenvectors with eigenvalue $+1$ and two with $-1$:
\begin{eqnarray}
\gamma_0 \xi_s = \xi_s \,, \qquad 
\gamma_0 \chi_s = -\chi_s \,.
\label{xichidef}
\end{eqnarray}
The subscripts on $\xi$ and $\chi$ distinguishes two different
eigenvectors of each kind.  Of course this guarantees that
\begin{eqnarray}
\xi_s^\dagger \chi_{s'} = 0 \,,
\label{xi.chi}
\end{eqnarray}
since they belong to different eigenvalues.  But since the two $\xi$'s
are degenerate and so are the two $\chi$'s, there is some
arbitrariness in defining them even for a given form of the matrix
$\gamma_0$.  In order to remove the arbitrariness, let us note that
the matrices $\sigma_{ij}$, with both space indices, commute with
$\gamma_0$.  In particular, say,
\begin{eqnarray}
\Big[ \sigma_{12} , \gamma_0 \Big] = 0 \,.
\end{eqnarray}
Thus, we can choose the eigenstates of $\gamma_0$ such that they are
simultaneously eigenstates of $\sigma_{12}$.  From
\Eqs{acomm}{sigdef}, it is easy to see that
\begin{eqnarray}
\Big( \sigma_{12} \Big)^2 = 1 \,,
\end{eqnarray}
so that the eigenvalues of $\sigma_{12}$ are $\pm1$ as well.
Therefore, let us choose the eigenvectors of $\gamma_0$ such that 
\begin{eqnarray}
\sigma_{12} \xi_s = s \xi_s \,, \qquad \sigma_{12} \chi_s = s \chi_s
\,, 
\label{sigxichi}
\end{eqnarray}
with $s=\pm$\,.  Once we fix the spinors in this manner, the four
eigenvectors are mutually orthogonal, i.e., in addition to
\Eqn{xi.chi}, the following relations also hold:
\begin{eqnarray}
\xi_s^\dagger \xi_{s'} = \delta_{ss'} \,, \qquad
\chi_s^\dagger \chi_{s'} = \delta_{ss'} \,.
\label{xichinorm}
\end{eqnarray}

One might wonder, why are we spending so much time in discussing the
eigenvectors of $\gamma_0$?  To see the reason, let us consider
\Eqn{uveqn} for vanishing 3-momentum.  In this case $E_{\vec p}=m$, so
that the equations reduce to
\begin{subequations}
\label{uv0eqn}
\begin{eqnarray}
( \gamma_0 - 1 ) u\((\vec 0\)) &=& 0 \,, \\ 
( \gamma_0 + 1 ) v\((\vec 0\)) &=& 0 \,.
\end{eqnarray}
\end{subequations}
This shows that, at zero momentum, the $u$-spinors and the $v$-spinors
are simply eigenstates of $\gamma_0$ with eigenvalues $+1$ and $-1$.
Thus we can define the zero-momentum spinors as
\begin{eqnarray}
u_s\((\vec 0\)) \propto \xi_s \,,\qquad 
v_s\((\vec 0\)) \propto \chi_{-s} \,,
\label{uv-}
\end{eqnarray}
apart from possible normalizing factors which will be specified later.

\subsection{Spinors and their normalization}
We now want to find the spinors for any value of $\vec p$.  We know
that these will have to satisfy \Eqs{ueqn}{veqn}, and, for $\vec p =
0$, should have the forms given in \Eqn{uv-}.  With these
observations, we can try the following solutions:
\begin{subequations}
\label{uvsoln}
\begin{eqnarray}
u_s\((\vec p\)) &=& N_{\vec p} (\gamma_\mu p^\mu + m) \xi_s 
\label{upxi} \,, \\ 
v_s\((\vec p\)) &=& N_{\vec p} (-\gamma_\mu p^\mu + m) \chi_{-s}  
\label{vpchi} \,,
\end{eqnarray}
\end{subequations}
where $N_{\vec p}$ is a normalizing factor.  One might wonder why we
have put $\chi_{-s}$ and not $\chi_s$ in the definition of $v_s$.  It
is nothing more than a convention.  It turns out that when we do
quantum field theory, this convention leads to an easy interpretation
of the subscript $s$.  This issue will not be discussed here.

It is easy to see that our choices for the spinors satisfy
\Eqn{uveqn} since
\begin{eqnarray}
(\gamma_\mu p^\mu - m) (\gamma_\nu p^\nu + m) = p^2 - m^2 = 0 \,.
\label{psq-msq}
\end{eqnarray}
It is also easy to see that in the zero-momentum limit, these
solutions reduce to the eigenvalues of $\gamma_0$, apart from a
normalizing factor.  For example, putting $\vec p = 0$ and $E_{\vec
p}=m$ into \Eqn{upxi}, we obtain
\begin{eqnarray}
u_s\((\vec 0\)) = N_{\vec 0} m (\gamma_0 + 1) \xi_s = 2m N_{\vec 0}
\xi_s \,.
\end{eqnarray}

In order to determine a convenient normalization of the spinors, let
us rewrite \Eqn{upxi} more explicitly:
\begin{eqnarray}
u_s\((\vec p\)) = N_{\vec p} (\gamma_0 E_{\vec p} - \gamma_i p_i + m)
\xi_s = N_{\vec p} ( E_{\vec p} + m - \gamma_i p_i) \xi_s \,,
\label{u}
\end{eqnarray}
using \Eqn{xichidef} in the last step.  Similarly, we obtain
\begin{eqnarray}
v_s\((\vec p\)) = N_{\vec p} ( E_{\vec p} + m + \gamma_i p_i) \chi_{-s}
\,.  
\label{v}
\end{eqnarray}
Recalling that $\gamma_i$'s are anti-hermitian matrices, we then
obtain 
\begin{subequations}
\label{uvdag}
\begin{eqnarray}
u_s^\dagger \((\vec p\)) &=& N_{\vec p}^* \xi_s^\dagger ( E_{\vec p} + m
+ \gamma_i p_i) \,, 
\label{udag}\\ 
v_s^\dagger \((\vec p\)) &=& N_{\vec p}^* \chi_{-s}^\dagger ( E_{\vec p}
+ m - \gamma_i p_i) \,.
\end{eqnarray}
\end{subequations}
Thus, 
\begin{eqnarray}
u_s^\dagger \((\vec p\)) u_{s'}\((\vec p\)) = \Big| N_{\vec p} \Big|^2
\xi_s^\dagger \Big( 
(E_{\vec p} + m)^2 - \gamma_i \gamma_j p_i p_j \Big) \xi_{s'} \,.
\end{eqnarray}
Since $p_i p_j=p_j p_i$, we can write
\begin{eqnarray}
\gamma_i \gamma_j p_i p_j = \frac12 \Big[ \gamma_i, \gamma_j \Big]_+
p_i p_j = - \delta_{ij} p_i p_j = - \vec p^2 \,.
\end{eqnarray}
Using \Eqn{Ep} then, we obtain
\begin{eqnarray}
u_s^\dagger \((\vec p\)) u_{s'}\((\vec p\)) = 2E_{\vec p} (E_{\vec p} +
m) \Big| N_{\vec p} \Big|^2 \xi_s^\dagger \xi_{s'} \,.
\end{eqnarray}
Choosing
\begin{eqnarray}
N_{\vec p} = {1 \over \sqrt{E_{\vec p} + m}} 
\label{Np}
\end{eqnarray}
and using \Eqn{xichinorm}, we obtain the normalization conditions in
the form
\begin{eqnarray}
u_s^\dagger \((\vec p\)) u_{s'}\((\vec p\)) = 2E_{\vec p}
\delta_{ss'} \,. 
\label{udagu}
\end{eqnarray}
Through a similar procedure, one can obtain a similar condition on the
$v$-spinors:
\begin{eqnarray}
v_s^\dagger \((\vec p\)) v_{s'}\((\vec p\)) = 2E_{\vec p} \delta_{ss'} \,.
\label{vdagv}
\end{eqnarray}

We now need a relation that expresses the orthogonality between a
$u$-spinor and a $v$-spinor.  In obtaining \Eqs{udagu}{vdagv}, the
linear terms in $\gamma_ip_i$, appearing in \Eqs{u}{udag} or in the
similar set of equations involving the $v$-spinors, cancel.  The same
will not work in combinations of the form $u_s^\dagger \((\vec p\))
v_{s'}\((\vec p\))$ because the $\gamma_ip_i$ terms have the same sign
in both factors.  However we notice that if we reverse the 3-momentum
in one of the factors, these problematic terms cancel.  We can then
follow the same steps, more or less, and use \Eqn{xi.chi} to obtain
\begin{eqnarray}
u_s^\dagger \((-\vec p\)) v_{s'}\((\vec p\)) = v_s^\dagger \((-\vec p\))
u_{s'}\((\vec p\)) = 0 \,.
\label{udagv}
\end{eqnarray}

\Eqn{udagv} can be expressed in an alternative form by using bars
rather than daggers, where $\bar w = w^\dagger \gamma_0$ for any
spinor.  Multiplying
\Eqn{ueqn} from the left by $\bar v_{s'}\(( \vec p \))$ we obtain
\begin{eqnarray}
\bar v_{s'}\(( \vec p \)) ( \gamma_\mu p^\mu - m ) u_s\((\vec p\)) =
0 \,. 
\end{eqnarray}
Multiplying the hermitian conjugate of the equation for $v_{s'}\((
\vec p \))$ by $u_s\((\vec p\))$ from the right, we get
\begin{eqnarray}
\bar v_{s'}\(( \vec p \)) ( \gamma_\mu p^\mu + m ) u_s\((\vec p\)) = 0
\,. 
\end{eqnarray}
Subtracting one of these equations from another, we find that
\begin{eqnarray}
\bar v_{s'}\(( \vec p \)) u_s\((\vec p\)) &=& 0 
\label{vbaru} 
\end{eqnarray}
provided $m\neq 0$.  Similarly, one can also obtain the equation
\begin{eqnarray}
\bar u_{s'}\(( \vec p \)) v_s\((\vec p\)) &=& 0 \,.
\label{ubarv}
\end{eqnarray}

We will also show, in \sec{s:blid}, that \Eqs{udagu}{vdagv}
are equivalent to the relations
\begin{subequations}
\label{barnorm}
\begin{eqnarray}
\bar u_s \((\vec p\)) u_{s'}\((\vec p\)) &=& 2m \delta_{ss'} \,,
\label{ubaru} \\
\bar v_s \((\vec p\)) v_{s'}\((\vec p\)) &=& - 2m \delta_{ss'} \,.
\label{vbarv}
\end{eqnarray}
\end{subequations}
Unless $m=0$, these can be taken as the normalization conditions on
the spinors.

\subsection{Spin sums}
The spinors also satisfy some completeness relations, which can be
proved without invoking their explicit forms \cite{LPbook}.  Consider
the sum
\begin{eqnarray}
A_u \((\vec p\)) \equiv \sum_s u_s\((\vec p\)) \bar u_s\((\vec p\))
\,. 
\end{eqnarray}
Note that, using \Eqn{ubaru}, we get 
\begin{eqnarray}
A_u \((\vec p\)) u_{s'}\((\vec p\)) &=& \sum_s u_s\((\vec p\)) \Big [\bar
u_s\((\vec p\)) u_{s'}\((\vec p\)) \Big] \nonumber\\ 
&=& 2m u_{s'}\((\vec p\)) \,.
\end{eqnarray}
And, using \Eqn{ubarv}, we get
\begin{eqnarray}
A_u \((\vec p\)) v_{s'}\((\vec p\)) = 0 \,.
\end{eqnarray}

Recalling \Eqs{ueqn}{veqn}, it is obvious that on the spinors
$u_s\((\vec p\))$ and $v_s\((\vec p\))$, the operation of $A_u \((\vec
p\))$ produces the same result as the operation of $\gamma_\mu
p^\mu+m$.  Since any 4-component column vector can be written as a
linear superposition of the basis spinors $u_s\((\vec p\))$ and
$v_s\((\vec p\))$, it means that the action of $A_u \((\vec p\))$ and
of $\gamma_\mu p^\mu+m$ produces identical results on any 4-component
column vector.  The two matrices must therefore be the same:
\begin{eqnarray}
\sum_s u_s\((\vec p\)) \bar u_s\((\vec p\)) = \gamma_\mu p^\mu+m \,.
\label{uubar}
\end{eqnarray}
Similar reasoning gives
\begin{eqnarray}
\sum_s v_s\((\vec p\)) \bar v_s\((\vec p\)) = \gamma_\mu p^\mu-m \,.
\label{vvbar}
\end{eqnarray}
%

\section{Relations involving spinors}\label{s:rl}
We now show some non-trivial properties of the spinors.  In all
textbooks, they are deduced in the Dirac-Pauli representation of the
$\gamma$-matrices.  Using \Eqn{gamtil}, one can show that if they hold
in one representation, they must hold in other representations as
well.  Here we derive them without using any representation at any
stage of the proofs.

\subsection{What $\gamma_0$ does on spinors}
We first consider the effect of $\gamma_0$ acting on the spinors.
From \Eqn{u}, we find
\begin{eqnarray}
\gamma_0 u_s\((\vec p\)) &=& N_{\vec p} \gamma_0 ( E_{\vec p} + m -
\gamma_i p_i) \xi_s \nonumber\\*
&=& N_{\vec p} ( E_{\vec p} + m + \gamma_i p_i) \gamma_0 \xi_s 
= N_{\vec p} ( E_{\vec p} + m + \gamma_i p_i) \xi_s \,,
\end{eqnarray}
using the anticommutation relations and \Eqn{xichidef}.  This shows
that 
\begin{eqnarray}
\gamma_0 u_s\((\vec p\)) = u_s\((-\vec p\)) \,.
\label{g0u}
\end{eqnarray}
Following the same procedure, we can obtain the result
\begin{eqnarray}
\gamma_0 v_s\((\vec p\)) = - v_s\((-\vec p\)) \,.
\label{g0v}
\end{eqnarray}
\Eqs{g0u}{g0v} are very important relations for deducing behavior of
fermions under the parity transformation.  These relations can be used
to deduce \Eqs{vbaru}{ubarv} from \Eqn{udagv}, or vice versa.

\subsection{Conjugation relations}
Let us now deduce another set of relations, which plays an important
role in deriving charge conjugation properties of fermions.  To build
up to these relations, let us first consider the object
\begin{eqnarray}
\hat\xi_s = \gamma_0 C \xi_s^* \,,
\end{eqnarray}
where the matrix $C$ was defined in \Eqn{C}.  To find out about the
nature of $\hat\xi_s$, we first consider the action of $\gamma_0$ on
it: 
\begin{eqnarray}
\gamma_0 \hat\xi_s = \gamma_0 \gamma_0 C \xi_s^* = - \gamma_0 C
\gamma_0^\top \xi_s^* \,,
\end{eqnarray}
using \Eqn{C} again.  However, the complex conjugate of \Eqn{xichidef}
implies that
\begin{eqnarray}
\gamma_0^\top \xi_s^* = \xi_s^* \,,
\end{eqnarray}
since 
\begin{eqnarray}
\gamma_0^* = \gamma_0^\top
\label{gam0*}
\end{eqnarray}
because of the hermiticity of the matrix $\gamma_0$.  Putting this in,
we obtain
\begin{eqnarray}
\gamma_0 \hat\xi_s = - \gamma_0 C \xi_s^* = - \hat\xi_s \,,
\end{eqnarray}
showing that $\hat\xi_s$ is an eigenvector of $\gamma_0$ with
eigenvalue $-1$.  Therefore, it must be a combination of the
$\chi_s$'s.  

To determine which combination of the $\chi_s$'s occur in $\hat\xi_s$,
we use \Eqn{CsigCinv} and recall that $\sigma_{12}$ commutes with
$\gamma_0$ to obtain
\begin{eqnarray}
\sigma_{12} \hat\xi_s = \gamma_0 \sigma_{12} C \xi_s^* = - \gamma_0 C 
\sigma_{12}^\top \xi_s^* \,.
\end{eqnarray}
It can be easily seen from \Eqs{gamdag}{sigdef} that $\sigma_{12}$ is
hermitian.  So, from \Eqn{sigxichi}, we obtain
\begin{eqnarray}
\sigma_{12}^\top \xi_s^* = \Big( \sigma_{12} \xi_s \Big)^* = s \xi_s^* \,,
\end{eqnarray}
which gives
\begin{eqnarray}
\sigma_{12} \hat\xi_s = - s \gamma_0 C \xi_s^* = -s \hat\xi_s \,.
\end{eqnarray}
This shows that $\hat\xi_s$ is also an eigenstate of $\sigma_{12}$,
with eigenvalue $-s$.  Recalling the result we found earlier about its
eigenvalue of $\gamma_0$, we conclude that $\hat\xi_s$ must be
proportional to $\chi_{-s}$.  Since both $\gamma_0$
and $C$ are unitary matrices and $\xi_s$ is normalized to have unit
norm, the norm of $\hat\xi_s$ is also unity, so the proportionality
constant can be a pure phase, of the form $e^{i\theta}$.  But notice
that the definition of the matrix $C$ in \Eqn{C} has a phase
arbitariness as well.  In other words, given a set of matrices
$\gamma_\mu$, the matrix $C$ can be obtained only up to an overall
phase from \Eqn{C}.  We can choose the overall phase of $C$ such that
the relation 
\begin{eqnarray}
\gamma_0 C \xi_s^* &=& \chi_{-s} 
\label{gam0Cxi*} 
\end{eqnarray}
is obeyed.  One can then see that 
\begin{eqnarray}
\gamma_0 C \chi_s^* &=& \gamma_0 C \Big( \gamma_0 C \xi_{-s}^* \Big)^* 
= \gamma_0 C \gamma_0^\top (C^\top)^\dagger \xi_{-s} \,,
\end{eqnarray}
using \Eqn{gam0*}.  At this stage, using \Eqs{C}{CT=-C}, we can write
\begin{eqnarray}
\gamma_0 C \chi_s^* &=& \gamma_0 \gamma_0 C C^\dagger \xi_{-s} \,.
\end{eqnarray}
Since $C$ is unitary and $\gamma_0$ squares to the unit matrix, we
obtain 
\begin{eqnarray}
\gamma_0 C \chi_s^* = \xi_{-s} \,,
\label{gam0Cchi*}
\end{eqnarray}
similar to \Eqn{gam0Cxi*}.

To see the implication of these relations between the eigenvectors of
$\gamma_0$, we take the complex conjugate of \Eqn{u}.  Remembering
that the matrices $\gamma_i$ are anti-hermititan so that
$\gamma_i^*=-\gamma_i^\top$, we obtain
\begin{eqnarray}
u_s^* \((\vec p\)) 
= N_{\vec p} ( E_{\vec p} + m + \gamma_i^\top p_i) \xi_s^* 
= N_{\vec p} ( E_{\vec p} + m - C^{-1} \gamma_i C p_i) \xi_s^* \,,
\label{u*}
\end{eqnarray}
using the definition of the matrix $C$ from \Eqn{C}.  Multiplying from
the left by $\gamma_0C$, we obtain
\begin{eqnarray}
\gamma_0C u_s^* \((\vec p\)) &=& N_{\vec p} \Big[ (E_{\vec p} + m)
  \gamma_0C \xi_s^*  - \gamma_0\gamma_i C p_i \xi_s^* \Big] \,.
\end{eqnarray}
Since $\gamma_0$ anticommutes with $\gamma_i$, this can be written as 
\begin{eqnarray}
\gamma_0C u_s^* \((\vec p\)) 
&=& N_{\vec p} \Big[ (E_{\vec p} + m)
   + \gamma_i p_i \Big] \gamma_0C \xi_s^* = N_{\vec p} \Big[ (E_{\vec
    p} + m) + \gamma_i p_i \Big] \chi_{-s} \,.
\end{eqnarray}
Using \Eqn{v}, we now obtain
\begin{eqnarray}
\gamma_0C u_s^* \((\vec p\)) = v_s\((\vec p\)) \,.
\label{gam0Cu*}
\end{eqnarray}
This is an important relation.  Following similar steps, we can also
prove the relation
\begin{eqnarray}
\gamma_0C v_s^* \((\vec p\)) = u_s\((\vec p\)) \,.
\label{gam0Cv*}
\end{eqnarray}
Because $C$ appears in the conjugation properties of the spinors, we
will sometimes refer to it as the conjugation matrix.

\subsection{What $\gamma_5$ does on spinors}
Multiplying both sides of \Eqn{ueqn} by $\gamma_5$ from the left and
using the anticommutation of $\gamma_5$ with all Dirac matrices, we
obtain the equation
\begin{eqnarray}
( \gamma_\mu p^\mu + m ) \gamma_5u\((\vec p\)) &=& 0 \,,
\end{eqnarray}
which clearly shows that $\gamma_5u$ is a $v$-spinor.  Similarly,
$\gamma_5v$ must be a $u$-spinor.  However, this simple argument does
not say whether $\gamma_5u_+$ is $v_+$, or $v_-$, or a linear
combination of the two.

To settle the issue, we note that
\begin{eqnarray}
\gamma_0 \gamma_5 \xi_s = - \gamma_5 \gamma_0 \xi_s = - \gamma_5
\xi_s \,,
\end{eqnarray}
since $\gamma_5$ anticommutes with $\gamma_0$.  This equation shows
that $\gamma_5\xi_s$ is an eigenvector of $\gamma_0$ with eigenvalue
$-1$, i.e., it must be some combination of the $\chi$-eigenvectors
defined in \Eqn{xichidef}.  Moreover, since $\gamma_5$ commutes with
$\sigma_{12}$, we observe that
\begin{eqnarray}
\sigma_{12} \gamma_5 \xi_s = \gamma_5 \sigma_{12} \xi_s = s \gamma_5
\xi_s \,.
\end{eqnarray}
This means that $\gamma_5 \xi_s$ is an eigenstate of $\sigma_{12}$
with eigenvalue $s$.  Combining this information about the eigenvalues
of $\gamma_0$ and $\sigma_{12}$, we conclude that $\gamma_5 \xi_s$
must be equal to $\chi_s$ apart from a possible constant phase factor.
Let us therefore write
\begin{eqnarray}
\gamma_5 \xi_s = \eta_s \chi_s \,,
\label{g5xi}
\end{eqnarray}
with $|\eta_s|=1$.  Because of \Eqn{g5sq}, this would also imply
\begin{eqnarray}
\gamma_5 \chi_s = \eta_s^* \xi_s \,.
\label{g5chi}
\end{eqnarray}

However, the two phase factors $\eta_s$ (for $s=\pm$) cannot be chosen
in a completely arbitrarily way, since we have utilized the freedom in
imposing \Eqs{gam0Cxi*}{gam0Cchi*}.  For example, we see that,
\begin{eqnarray}
\chi_{-s} &=& \gamma_0 C\xi_s^* = \gamma_0 C \Big( \eta_s \gamma_5
\chi_s \Big)^* = \eta_s^* \gamma_0 C \gamma_5^\top \chi_s^* \,,
\end{eqnarray}
using the hermiticity of the matrix $\gamma_5$.  Now, using
\Eqn{Cinvg5C}, we can further simplify this expression and write
\begin{eqnarray}
\chi_{-s} &=& \eta_s^* \gamma_0 \gamma_5 C \chi_s^*
= - \eta_s^* \gamma_5 \gamma_0 C \chi_s^* = - \eta_s^* \gamma_5
\xi_{-s} = - \eta_s^* \eta_{-s} \chi_{-s} \,,
\end{eqnarray}
using \Eqs{gam0Cchi*}{g5xi} on the way.  This means that, the choice
of phases implied in writing \Eqs{gam0Cxi*}{gam0Cchi*} forces us to
impose the relation
\begin{eqnarray}
\eta_s^* \eta_{-s} = - 1 \,.
\end{eqnarray}
One possible way of assuring this relation is to take
\begin{eqnarray}
\eta_s = (-1)^{(s-1)/2} \,,
\label{etachoice}
\end{eqnarray}
although we will not use the specific choice in what follows.

The action of $\gamma_5$ on the spinors can now be calculated easily.
For example, one finds
\begin{eqnarray}
\gamma_5 u_s\(( \vec p \)) &=& N_{\vec p} \gamma_5 (\gamma_\mu p^\mu + m)
\xi_s = N_{\vec p} (-\gamma_\mu p^\mu + m) \gamma_5 \xi_s \nonumber\\* 
&=& N_{\vec p} (-\gamma_\mu p^\mu + m) \eta_s \chi_s = \eta_s v_{-s}
\(( \vec p \)) \,. 
\label{g5u}
\end{eqnarray}
Through similar manipulations or through the use of \Eqn{g5sq}, we can
get 
\begin{eqnarray}
\gamma_5 v_s \(( \vec p \)) = - \eta_s^* u_{-s} \(( \vec p \)) \,. 
\label{g5v}
\end{eqnarray}
%

\subsection{Alternative forms}
The results obtained above can be combined to obtain some other
relations.  For example, multiply both sides of \Eqn{gam0Cu*} from the
left by $C^{-1}$.  Using \Eqn{C}, the result can be written as
\begin{eqnarray}
- \gamma_0^\top u_s^* \((\vec p\)) = C^{-1} v_s\((\vec p\)) \,.
\end{eqnarray}
But $\gamma_0^\top=\gamma_0^*$, so the left hand side is the complex
conjugate of $\gamma_0u_s \((\vec p\))$.  Using \Eqn{g0u}, we can then
write 
\begin{eqnarray}
C^{-1} v_s\((\vec p\)) = - u_s^* \((- \vec p\)) \,.
\end{eqnarray}
Similar manipulations give the complimentary result,
\begin{eqnarray}
C^{-1} u_s\((\vec p\)) = v_s^* \((- \vec p\)) \,.
\end{eqnarray}

We can also combine this result with the identities of \Eqs{g5u}{g5v}
to obtain
\begin{eqnarray}
C^{-1} \gamma_5 v_s\((\vec p\)) &=& - \eta_s^* v_{-s}^* \(( -\vec p \)) \,,
\nonumber\\* 
C^{-1} \gamma_5 u_s\((\vec p\)) &=& - \eta_s u_{-s}^* \(( -\vec p \)) \,.
\end{eqnarray}
The matrix $C^{-1} \gamma_5$ plays a crucial role in the time-reversal
properties of a fermion field~\cite{LPbook}.

\section{Spinor bilinears}\label{s:bl}
Whenever fermion fields have to be used in Lorentz invariant
combinations, we must encounter pairs of them in order that the
overall combination conserves angular momentum.  For this reason,
fermion field bilinears deserve some attention.  In momentum space,
one encounters bilinears involving spinors, which is what we discuss
in this section.

\subsection{Identities involving bilinears}\label{s:blid}
A vector $p_\lambda$ can be rewritten as
\begin{eqnarray}
p_\lambda = g_{\lambda\rho} p^\rho = \Big( \gamma_\lambda
\gamma_\rho + i \sigma_{\lambda\rho} \Big) p^\rho =  \gamma_\lambda
\rlap/ p + i \sigma_{\lambda\rho} p^\rho \,.
\end{eqnarray}
Alternatively, we can write
\begin{eqnarray}
p'_\lambda = g_{\lambda\rho} p'^\rho = \Big( \gamma_\rho
\gamma_\lambda + i \sigma_{\rho\lambda} \Big) p^\rho 
= \rlap/p' \gamma_\lambda - i \sigma_{\lambda\rho} p'^\rho \,. 
\end{eqnarray}
Adding these two equation, sandwiching the result between two spinors,
and using \Eqn{ueqn} and its hermitian conjugate, we obtain the
relation
\begin{eqnarray}
\bar u \((\vec p'\)) \gamma_\lambda u\((\vec p\)) = {1 \over 2m}
\bar u \((\vec p'\)) \Big[ Q_\lambda -i\sigma_{\lambda\rho}
  q^\rho \Big] u\((\vec p\)) \,,
\label{Gordon}
\end{eqnarray}
where
\begin{eqnarray}
Q = p + p' \,, \qquad  q = p - p' \,.
\label{Qq}
\end{eqnarray}
This result is called the Gordon identity.

Variants of this identity can be easily derived following the same
general technique.  For example, suppose the two spinors on the two
sides belong to different particles, with masses $m$ and $m'$.  In
this case, it is easy to see that
\begin{eqnarray}
\bar u \((\vec p'\)) \Big[ Q_\lambda -i\sigma_{\lambda\rho}
  q^\rho \Big] u\((\vec p\)) = (m'+m) \bar u \((\vec p'\))
  \gamma_\lambda u\((\vec p\)) \,.
\label{Gordonmm'}
\end{eqnarray}
Similarly, one can obtain the identity
\begin{eqnarray}
\bar u \((\vec p'\)) \Big[ Q_\lambda -i\sigma_{\lambda\rho}
  q^\rho \Big] \gamma_5 u\((\vec p\)) = (m'-m) \bar u \((\vec p'\))
  \gamma_\lambda \gamma_5 u\((\vec p\)) \,.
\label{Gordon5}
\end{eqnarray}

It should be noted that the normalization relations of
\Eqs{udagu}{vdagv} can be written in an alternative form by using the
Gordon identity.  For this, we put $\vec p = \vec p'$ in \Eqn{Gordon}
and take only the time component of the equation.  This gives
\begin{eqnarray}
2m \, u_{s'}^\dagger \((\vec p\)) u_s\((\vec p\)) = 2E_{\vec p}
\bar u_{s'}\((\vec p\)) u_s\((\vec p\)) \,,
\end{eqnarray}
where we have put the indices $s,s'$ on the spinors in order to
distinguish the different solutions.  This shows that \Eqs{udagu}{ubaru}
are equivalent.  The proof of the equivalence of \Eqs{vdagv}{vbarv} is
similar.

\subsection{Non-relativistic reduction}
In field-theoretical manipulations, sometimes we encounter expressions
which can be interpreted easily by making a non-relativistic
reduction.  For example, in Quantum Electrodynamics (QED), the matrix
element of the electromagnetic current operator turns out to be
superposition of two bilinears of the form $\bar u\((\vec p'\))
\gamma_\lambda u\((\vec p\))$ and $\bar u\((\vec p'\))
\sigma_{\lambda\rho} q^\rho u\((\vec p\))$, and an intuitive feeling
for these bilinears can be obtained by going to the non-relativistic
limit.  With this in mind, here we give the non-relativistic reduction
of all possible fermion bilinears.

A general bilinear is of the form
\begin{eqnarray}
\bar u_{s'}\((\vec p'\)) F u_s\((\vec p\))
\end{eqnarray}
for some matrix $F$.  Any such matrix can be written as a
superposition of the 16 basis matrices shown in \Eqn{basis16}.  So it
is enough to obtain non-relativistic reduction with the bilinears
involving these basis matrices only.

We will keep terms up to linear order in the 3-momenta.  The
spinor, to this order, can be written as
\begin{eqnarray}
u_s\((\vec p\)) \approx \sqrt{2m} \left( 1 - {\gamma_i p_i \over 2m}
\right) \xi_s \,, 
\label{u1}
\end{eqnarray}
using \Eqs{u}{Np}, where the `approximate equal to' sign ($\approx$)
will be used throughout this section to imply that all terms of the
order 3-momentum squared have been omitted.   Then
\begin{eqnarray}
\bar u_{s'}\((\vec p'\)) \approx \sqrt{2m} 
\xi_{s'}^\dagger \left( 1 - {\gamma_i p'_i \over 2m} \right) \,, 
\end{eqnarray}
using the hermiticity and anticommutation properties of the gamma
matrices, and the fact that
\begin{eqnarray}
\xi_{s'}^\dagger \gamma_0 = \xi_{s'}^\dagger 
\label{xidag0}
\end{eqnarray}
which follows from \Eqn{xichidef}.  For the general bilinear, then, we
obtain
\begin{eqnarray}
\bar u_{s'}\((\vec p'\)) F u_s\((\vec p\)) \approx \xi_{s'}^\dagger \Big( 
2mF - \frac12 Q_j \Big[ F, \gamma_j \Big]_+ - \frac12 q_j \Big[ F,
  \gamma_j \Big] \Big) \xi_s \,,
\label{upto1}
\end{eqnarray}
using the sum and difference of momenta introduced in \Eqn{Qq}.  The
three terms on the right side of this equation will be referred to as
the momentum-independent term, the anticommutator term and the
commutator term respectively.  We now evaluate these terms for the
five types of basis matrices shown in \Eqn{basis16}.

\subsubsection{Scalar bilinear}
This corresponds to the case $F=\umat$, so \Eqn{upto1} for this case
reads 
\begin{eqnarray}
\bar u_{s'}\((\vec p'\)) F u_s\((\vec p\)) \approx \xi_{s'}^\dagger
\Big( 2m \umat - Q_j \gamma_j \Big) \xi_s \,.
\end{eqnarray}
In the second term on the right side, one can use the definition of
$\xi_s$ from \Eqs{xichidef}{xidag0} as well as the anticommutation of
$\gamma_0$ with all $\gamma_i$'s to write
\begin{eqnarray}
\xi_{s'}^\dagger \gamma_i \xi_s = \xi_{s'}^\dagger \gamma_0 \gamma_i
\xi_s
= - \xi_{s'}^\dagger \gamma_i \gamma_0 \xi_s = - \xi_{s'}^\dagger
\gamma_i \xi_s \,,
\label{blgammai}
\end{eqnarray}
so that
\begin{eqnarray}
\xi_{s'}^\dagger \gamma_i \xi_s = 0 \,.
\label{xigamixi}
\end{eqnarray}
The momentum-independent term can be easily written down using
\Eqn{xichinorm}, and one obtains
\begin{eqnarray}
\bar u_{s'}\((\vec p'\)) u_s\((\vec p\)) \approx 2m \delta_{ss'} \,.
\end{eqnarray}
Recall that with $\vec p = \vec p'$, this is the equality of
\Eqn{ubaru}.  This equation shows that even if $\vec p \neq \vec p'$,
the corrections appear only in the second order of the momenta.

\subsubsection{Vector bilinears}
These corresponds to $F=\gamma_\lambda$ for some index $\lambda$.
Consider first the case when $\lambda$ is a spatial index.  Note that
the momentum-independent term for this case vanishes due to the
identity of \Eqn{xigamixi}.  Thus we obtain
\begin{eqnarray}
\bar u_{s'}\((\vec p'\)) \gamma_i u_s\((\vec p\)) \approx \xi_{s'}^\dagger
\Big[ \delta_{ij} Q_j + i \sigma_{ij} q_j \Big] \xi_s \,,
\end{eqnarray}
using the anticommutator and commutator of the gamma matrices from
\Eqs{acomm}{sigdef}.  In fact, this result follows directly from the
Gordon identity, \Eqn{Gordon}, if we keep only terms up to first order
in momenta.  Using the definition of the spin matrices, \Eqn{Sigma}, we
can also write the equation in the form
\begin{eqnarray}
\bar u_{s'}\((\vec p'\)) \gamma_i u_s\((\vec p\)) \approx 
Q_i \delta_{ss'} + i \varepsilon_{ijk} q_j \; \xi_{s'}^\dagger \Sigma_k
\xi_s \,. 
\end{eqnarray}

We now turn to the temporal part of the matrix element, i.e., the case
with $F=\gamma_0$ in \Eqn{upto1}.  Since $\gamma_0$ anticommutes with
all $\gamma_j$, we obtain
\begin{eqnarray}
\bar u_{s'}\((\vec p'\)) \gamma_0 u_s\((\vec p\)) \approx \xi_{s'}^\dagger 
\Big( 
2m \gamma_0 - \frac12 (\gamma_0 \gamma_j - \gamma_j \gamma_0) q_j
\Big) \xi_s \,. 
\end{eqnarray}
Using \Eqs{xichidef}{xidag0}, we find that the term linear in momenta
vanishes, so that, to the order specified, we obtain
\begin{eqnarray}
\bar u_{s'}\((\vec p'\)) \gamma_0 u_s\((\vec p\)) \approx 
2m \delta_{ss'} \,.
\end{eqnarray}
%

\subsubsection{Tensor bilinears}
For tensor bilinears, $F=\sigma_{\lambda\rho}$.  Individual terms that
appear in the commutator and anticommutator involving $F$ appearing in
\Eqn{upto1} are products of three gamma matrices.  All such terms can
be reduced by using \Eqn{ggg}, and one obtains
\begin{subequations}
\label{siggam}
\begin{eqnarray}
\Big[ \sigma_{\lambda\rho} , \gamma_\mu \Big] &=& 2i (g_{\mu\rho}
  \gamma_\lambda - g_{\mu\lambda} \gamma_\rho) \,, \\
\Big[ \sigma_{\lambda\rho} , \gamma_\mu \Big]_+ &=& 2
\varepsilon_{\lambda\rho\mu\nu} \gamma^\nu \gamma_5 \,.
\end{eqnarray}
\end{subequations}

In particular, if we consider the sigma matrices with one temporal
index, we need the relations
\begin{subequations}
\begin{eqnarray}
\Big[ \sigma_{0i} , \gamma_j \Big] &=& 2i g_{ji} \gamma_0 = -2i
\delta_{ij} \gamma_0 \,, \\
\Big[ \sigma_{0i} , \gamma_j \Big]_+ &=& 2
\varepsilon_{0ijk} \gamma^k \gamma_5 = - 2 \varepsilon_{ijk} \gamma_k
\gamma_5 = 2 \varepsilon_{ijk} \gamma_0 \Sigma_k \,,
\end{eqnarray}
\end{subequations}
using \Eqn{0i5} on the way.  Thus, from \Eqn{upto1}, we obtain
\begin{eqnarray}
\bar u_{s'}\((\vec p'\)) \sigma_{0i} u_s\((\vec p\)) 
\approx \xi_{s'}^\dagger \Big( 
2m \sigma_{0i} - \varepsilon_{ijk} Q_j \gamma_0 \Sigma_k 
+ iq_i \gamma_0 \Big) \xi_s \,. 
\end{eqnarray}
Using \Eqn{xichidef} throughout, we see that the momentum-independent
term on the right side vanishes, and we are left with 
\begin{eqnarray}
\bar u_{s'}\((\vec p'\)) \sigma_{0i} u_s\((\vec p\)) \approx 
i q_i \delta_{ss'} 
- Q_j \varepsilon_{ijk}  \xi_{s'}^\dagger \Sigma_k \xi_s \,.
\end{eqnarray}

A different kind of non-relativistic limit is obtained if both 
indices on the sigma-matrix are spatial.  The matrix $\sigma_{ij}$ is
essentially a spin matrix, as mentioned in \Eqn{Sigma}.  The
momentum-independent term in $\bar u_{s'}\(( \vec p'\)) \sigma_{ij} u_s\((
\vec p\))$ is then the matrix element of the spin operator.  In
particular, if the spinors on the two sides have $s=s'$, then the
bilinear is the expectation value of spin in that state.  In order to
evaluate the terms linear in the momenta, we need the following
relations which follow from \Eqn{siggam}:
\begin{subequations}
\begin{eqnarray}
\Big[ \sigma_{ij} , \gamma_k \Big] &=& 2i (g_{jk}
  \gamma_i - g_{ik} \gamma_j) \,, \\
\Big[ \sigma_{ij} , \gamma_k \Big]_+ &=& 2
\varepsilon_{ijk\nu} \gamma^\nu \gamma_5 = - 2 \varepsilon_{ijk}
\gamma_0 \gamma_5 \,. 
\end{eqnarray}
\end{subequations}
Obviously, the commutator term does not give a non-zero contribution
to the bilinear of \Eqn{upto1} because of \Eqn{xigamixi}.  Since
$\gamma_5$ anticommutes with $\gamma_0$, we can use the steps shown in
\Eqn{blgammai}, with $\gamma_i$ replaced by $\gamma_5$, to show that
\begin{eqnarray}
\xi_{s'}^\dagger \gamma_5 \xi_s = 0 \,.
\label{xigam5xi}
\end{eqnarray}
So even the anticommutator term does not contribute.  Only the
momentum-independent term survives to this order, and the result is
\begin{eqnarray}
\bar u_{s'}\((\vec p'\)) \sigma_{ij} u_s\((\vec p\)) \approx 
2m \varepsilon_{ijk} \; \xi_{s'}^\dagger \Sigma_k \xi_s \,.
\end{eqnarray}
%

\subsubsection{Pseudoscalar bilinear}
This corresponds to the case $F=\gamma_5$.   The momentum-independent
term vanishes because of \Eqn{xigam5xi}, and the anticommutator is
also zero, so that we are left with
\begin{eqnarray}
\bar u_{s'}\((\vec p'\)) \gamma_5 u_s\((\vec p\)) \approx 
q_j \xi_{s'}^\dagger \gamma_j \gamma_5 \xi_s \,,
\end{eqnarray}
Using \Eqn{0i5}, this expression can be written in the form
\begin{eqnarray}
\bar u_{s'}\((\vec p'\)) \gamma_5 u_s\((\vec p\)) \approx 
- q_j \xi_{s'}^\dagger \Sigma_j \xi_s \,,
\end{eqnarray}
recalling the definition of $\xi$ in \Eqn{xichidef}.

\subsubsection{Axial vector bilinears}
Finally, we discuss the cases when $F$ is of the form
$\gamma_\lambda\gamma_5$.  Two different cases arise, as in the case
with vector or tensor bilinears.  For $F=\gamma_0\gamma_5$, we can use
\Eqs{xidag0}{xigam5xi} to write 
\begin{eqnarray}
\xi_{s'}^\dagger \gamma_0 \gamma_5 \xi_s = \xi_{s'}^\dagger \gamma_5
\xi_s = 0 \,, 
\end{eqnarray}
which means that the momentum-independent term vanishes.  The
commutator term is also zero, so that we are left with
\begin{eqnarray}
\bar u_{s'}\((\vec p'\)) \gamma_0 \gamma_5 u_s\((\vec p\)) 
\approx - Q_j \xi_{s'}^\dagger 
\gamma_0 \gamma_5 \gamma_j \xi_s \,,
\end{eqnarray}
Using \Eqn{0i5} now, this can be written as
\begin{eqnarray}
\bar u_{s'}\((\vec p'\)) \gamma_0 \gamma_5 u_s\((\vec p\)) 
\approx - Q_i \xi_{s'}^\dagger \Sigma_i  \xi_s \,,
\end{eqnarray}

On the other hand, for $F=\gamma_i\gamma_5$, we find that the
commutator appearing in \Eqn{upto1} is
\begin{eqnarray}
\Big[ \gamma_i \gamma_5, \gamma_j \Big] = - \Big[ \gamma_i, \gamma_j
  \Big]_+ \gamma_5 = 2 \delta_{ij} \gamma_5 \,,
\end{eqnarray}
whose matrix element vanishes because of \Eqn{xigam5xi}.  The
anticommutator is
\begin{eqnarray}
\Big[ \gamma_i \gamma_5, \gamma_j \Big]_+ = - 2i \sigma_{ij} \gamma_5 \,.
\end{eqnarray}
However, using \Eqs{Sigma}{0i5}, we find
\begin{eqnarray}
\sigma_{ij} \gamma_5 = \varepsilon_{ijk} \Sigma_k \gamma_5 =
\varepsilon_{ijk} \gamma_0 \gamma_k \,,
\end{eqnarray}
whose matrix element also vanishes owing to \Eqs{xidag0}{xigamixi}.
Thus, only the momentum-independent term survives to the proposed
order, and the result can be written as
\begin{eqnarray}
\bar u_{s'}\((\vec p'\)) \gamma_i \gamma_5 u_s\((\vec p\)) 
\approx - \xi_{s'}^\dagger \Sigma_i \xi_s \,,
\end{eqnarray}
using \Eqs{0i5}{xidag0}.  Once again, these are matrix elements of the
spin operator, which reduce to the expectation value of spin if the
two spinors on both sides are the same.

\subsubsection{A note on the momentum expansion of bilinears}
We see that, in the momentum expansion, bilinears which have a
momentum-independent term do not have a term that is linear in the
3-momenta, and vice versa.  This feature can be explained by using
the parity properties of the bilinears.  Here we outline a proof
without invoking parity explicitly.

For this, consider two different representations of the Dirac
matrices, one denoted by a tilde sign and one without, which are
related in the following way:
\begin{eqnarray}
\tilde\gamma_0 = \gamma_0 \,, \qquad 
\tilde\gamma_i = -\gamma_i \,. 
\end{eqnarray}
Obviously, if the $\gamma_\mu$'s satisfy the anticommutation relation,
so do the $\tilde\gamma_\mu$'s.  The eigenvectors $\xi_s$ will be
identical in the two representations, since $\gamma_0$ is the same.
From \Eqn{u}, we see that the spinors in the tilded representation
are given by
\begin{eqnarray}
\tilde u_s\((-\vec p\)) = u_s\((\vec p\)) \,.
\label{util}
\end{eqnarray}

The bilinears are representation-independent.  Thus, 
\begin{eqnarray}
\bar{\tilde u}_{s'}\((\vec p'\)) \tilde F \tilde u_s\((\vec p\))
= \bar u_{s'}\((\vec p'\)) F u_s\((\vec p\)) \,,
\end{eqnarray}
where $\tilde F$ contains exactly the same string of Dirac matrices or
associated matrices of the tilded representation that are contained in
$F$, e.g., if $F=\sigma_{0i}$ then $\tilde F = \tilde\sigma_{0i}$.
Using \Eqn{util} now, we can write
\begin{eqnarray}
\bar u_{s'}\((-\vec p'\)) \tilde F u_s\((-\vec p\)) 
= \bar u_{s'}\((\vec p'\)) F u_s\((\vec p\)) \,,
\end{eqnarray}
If $F$ contains an even number of $\gamma_i$'s, then $F$ and $\tilde
F$ are equal, and we see that the bilinear would contain only even
order terms in the 3-momenta.  This is the case if $F$ is $\umat$,
$\gamma_0$, $\sigma_{ij}$ or $\gamma_i\gamma_5$.  Note that the
definition of \Eqn{g5def} implies that $\gamma_5$ contains an odd
number of $\gamma_i$'s.  On the other hand, if $F$ is any of the
combinations $\gamma_i$, $\gamma_0\gamma_5$, $\sigma_{0i}$ and
$\gamma_5$, the bilinears are odd in the 3-momenta.  We have seen 
these features explicitly in the reductions of the bilinears above.

\section{Spinor quadrilinears: Fierz identities}\label{s:fi}
A quadrilinear is a product of two bilinears of spinors.  This kind of
objects appear in the low-energy limit of any theory where fermions
interact through exchanges of bosons, e.g., in the Fermi theory of
weak interactions.  The important point is that there is some
arbitrariness in the order of the spinors in writing quadrilinears,
expressed through identities which are called Fierz identities
\cite{fierz}.  This is the subject of discussion of this section.

We will denote spinors by $w_1$, $w_2$ etc.\ in this section.  Here,
the letter $w$ can stand for either $u$ or $v$, i.e., each of the
spinors that appear in this section can be either a $u$-spinor or a
$v$-spinor.  The subscript 1,2 etc.\ stand for a certain 3-momentum
and a certain mass.  For example, $w_1$ can mean a $u$-spinor with
momentum $\vec p_1$ for a particle of mass $m_1$, and so on.

In order to pave the road for the Fierz identities, we first consider
a product $w_2 \bar w_1$.  It is a $4 \times 4$ matrix, and therefore
can be written in the form given in \Eqn{M}:
\begin{eqnarray}
w_2 \bar w_1 = a + b^\mu \gamma_\mu + c^{\mu\nu} \sigma_{\mu\nu} 
+ d^\mu \gamma_\mu\gamma_5 + e \gamma_5 \,.
\label{w2w1}
\end{eqnarray}
In order to evaluate the co-efficients $a$ through $e$, we first take
the trace of this expression.  Using the facts that the trace of any
odd number of $\gamma$-matrices is zero, and the trace of
$\sigma_{\mu\nu}$ is zero because it is a commutator, and
\Eqn{trgam5}, we obtain
\begin{eqnarray}
\tr \Big( w_2 \bar w_1 \Big) = a \tr \umat = 4a \,.
\end{eqnarray}
Since the trace operation is cyclic, we can write this equation as
\begin{eqnarray}
a = \frac14 \tr \Big( \bar w_1 w_2 \Big) = \frac14 \bar w_1 w_2 \,,
\end{eqnarray}
using in the last step the fact that $\bar w_1w_2$, being just a
$1\times 1$ matrix, is the trace of itself.  Exactly similarly,
multiplying both sides of \Eqn{w2w1} by $\gamma_\lambda$ from the left
before taking the trace, we would obtain
\begin{eqnarray}
b^\mu = \frac14 \Big( \bar w_1 \gamma^\mu w_2 \Big) \,.
\end{eqnarray}
We can continue this process to evaluate all co-efficients of
\Eqn{w2w1}, and the result is
\begin{eqnarray}
w_2 \bar w_1 &=& \frac14 \Big[  
\Big( \bar w_1 w_2 \Big) \umat + 
\Big( \bar w_1 \gamma^\mu w_2 \Big) \gamma_\mu + 
\frac12 \Big( \bar w_1 \sigma^{\mu\nu} w_2 \Big) \sigma_{\mu\nu} 
\nonumber\\*
&& \qquad -
\Big( \bar w_1 \gamma^\mu \gamma_5 w_2 \Big) \gamma_\mu\gamma_5 +
\Big( \bar w_1 \gamma_5 w_2 \Big) \gamma_5 \Big] \,.
\label{basicFI}
\end{eqnarray}
This is the basic Fierz identity.  The identities involving
quadrilinears follow from it.  For example, one can multiply both
sides by $\bar w_3$ from the left and by $w_4$ from the right,
obtaining 
\begin{eqnarray}
\Big( \bar w_3 w_2 \Big) \Big( \bar w_1 w_4 \Big) &=& \frac14 \Big[  
\Big( \bar w_1 w_2 \Big) \Big( \bar w_3 w_4 \Big) + \Big( \bar w_1
\gamma^\mu w_2 \Big) \Big( \bar w_3 \gamma_\mu w_4 \Big) \nonumber\\*  
&& \qquad + 
\frac12 \Big( \bar w_1 \sigma^{\mu\nu} w_2 \Big) \Big( \bar w_3
\sigma_{\mu\nu} w_4 \Big) 
- \Big( \bar w_1 \gamma^\mu \gamma_5 w_2 \Big) \Big( \bar w_3
\gamma_\mu\gamma_5 w_4 \Big) \nonumber\\*
&& \qquad + 
\Big( \bar w_1 \gamma_5 w_2 \Big) \Big( \bar w_3 \gamma_5 w_4 \Big)
\Big] \,. 
\label{scalarFI}
\end{eqnarray}
Similarly, if one multiplies \Eqn{basicFI} by $\bar w_3
\gamma^\lambda$ from the left and by $\gamma_\lambda w_4$ from the
right, one obtains $\Big( \bar w_3 \gamma^\lambda w_2 \Big) \Big( \bar
w_1 \gamma_\lambda w_4 \Big)$ on the left side.  On the right side,
the bilinears of the form $\bar w_3 \cdots w_4$ that appear are the
following: 
\begin{subequations}
\begin{eqnarray}
\bar w_3 \gamma^\lambda \gamma_\lambda w_4 &=& 4 \bar w_3 w_4 \,, \\
\bar w_3 \gamma^\lambda \gamma_\mu \gamma_\lambda w_4 &=& -2 \bar w_3
\gamma_\mu w_4 \,, \\*  
\bar w_3 \gamma^\lambda
\sigma_{\mu\nu} \gamma_\lambda w_4  &=& 0 \,, \\
\bar w_3 \gamma^\lambda \gamma_\mu\gamma_5 \gamma_\lambda w_4 &=& 2
\bar w_3 \gamma_\mu \gamma_5 w_4 \,,
\\* 
\bar w_3 \gamma^\lambda \gamma_5 \gamma_\lambda w_4 &=& -4 \bar w_3
\gamma_5 w_4 \,, 
\end{eqnarray}
\end{subequations}
where various contraction formulas listed in \Eqn{contr} and
\Eqn{gamsiggam} have been used to simplify the left sides of these
equations.  Thus, the final form of this Fierz identity would be
\begin{eqnarray}
\Big( \bar w_3 \gamma^\lambda w_2 \Big) \Big( \bar w_1 \gamma_\lambda
w_4 \Big) &=& \frac14 \Big[ 4 
\Big( \bar w_1 w_2 \Big) \Big( \bar w_3 w_4 \Big) -2 \Big( \bar w_1
\gamma^\mu w_2 \Big) \Big( \bar w_3 \gamma_\mu w_4 \Big) \nonumber\\*  
&& \quad 
-2 \Big( \bar w_1 \gamma^\mu \gamma_5 w_2 \Big) \Big( \bar w_3 
\gamma_\mu\gamma_5 w_4 \Big) - 4
\Big( \bar w_1 \gamma_5 w_2 \Big) \Big( \bar w_3 \gamma_5 w_4 \Big)
\Big] \,.
\nonumber\\* 
\end{eqnarray}
Identities involving other kinds of bilinears on the left side can be
easily constructed.

\section{Concluding remarks}\label{s:cr}
The aim of the article was to show that some important identities
involving Dirac spinors can be proved without invoking any specific
form for the spinors.  As we mentioned earlier, the specific forms
depend on the representation of the Dirac matrices.  For the sake of
elegance and safety, it is better to deal with the spinors in a
representation-independent manner.

The analysis can be extended to quantum field theory involving Dirac
fields.  Properties of Dirac field under parity, charge conjugation
and time reversal can be derived in completely
representation-independent manner.  This has been done at least in one
textbook of quantum field theory \cite{LPbook}, to which we refer the
reader for details.

\paragraph*{Acknowledgements~:}
I am indebted to E. Akhmedov for pointing out an
inconsistency regarding the choice of phases of spinors that appeared
in an earlier version of the paper and suggesting alternatives to
avoid the problem.  His help was also crucial for
Sec.\,\ref{ss:Ctrans}, as described in Ref.\,\cite{Ctrans}.

\appendix
\let\theSecTion=\thesection
\renewcommand\theequation{\theSecTion.\arabic{equation}}
\renewcommand\thesection{Appendix~\Alph{section}}
\section{The metric tensor and the Levi-Civita symbol}\label{s:id}
Our convention for the metric tensor is:
\begin{eqnarray}
g_{\mu\nu} = \mathop{\rm diag} (+1,-1,-1,-1) \,.
\label{g}
\end{eqnarray}
The Levi-Civita symbol is the completely antisymmetric rank-4 tensor,
whose non-zero elements have been chosen by the convention given in
\Eqn{LCconv}.  Product of two Levi-Civita symbols can be expressed in
terms of the metric tensor:
\begin{eqnarray}
\varepsilon^{\mu\nu\lambda\rho} \, \varepsilon_{\mu'\nu'\lambda'\rho'} =
- \left\Vert
  \begin{array}{c@{\hspace{2mm}}c@{\hspace{2mm}}c@{\hspace{2mm}}c}
    \delta^\mu_{\mu'} & \delta^\mu_{\nu'} & \delta^\mu_{\lambda'} &
    \delta^\mu_{\rho'} \\ 
    \delta^\nu_{\mu'} & \delta^\nu_{\nu'} & \delta^\nu_{\lambda'} &
    \delta^\nu_{\rho'} \\ 
    \delta^\lambda_{\mu'} & \delta^\lambda_{\nu'} &
    \delta^\lambda_{\lambda'} & \delta^\lambda_{\rho'} \\ 
    \delta^\rho_{\mu'} & \delta^\rho_{\nu'} & \delta^\rho_{\lambda'} &
    \delta^\rho_{\rho'}
  \end{array}
\right\Vert 
\equiv - \delta^\mu_{[\mu'} \delta^\nu_{\nu'}
\delta^\lambda_{\lambda'} \delta^\rho_{\rho']} \,,
\label{epseps}
\end{eqnarray}
where the pair of two vertical lines on two sides of the matrix
indicates the determinant of the matrix, and the square brackets
appearing among the indices imply an antisymmetrization with respect
to the enclosed indices.  By taking successive contractions of this
relation, we can obtain the following relations:
\begin{subequations}
\label{epseps1234}
\begin{eqnarray}
\varepsilon^{\mu\nu\lambda\rho} \, \varepsilon_{\mu\nu'\lambda'\rho'} 
&=& - \delta^\nu_{[\nu'}
\delta^\lambda_{\lambda'} \delta^\rho_{\rho']} \,, 
\label{epseps1} \\
\varepsilon^{\mu\nu\lambda\rho} \, \varepsilon_{\mu\nu\lambda'\rho'} 
&=& - 2 
\delta^\lambda_{[\lambda'} \delta^\rho_{\rho']} \,, 
\label{epseps2} \\
\varepsilon^{\mu\nu\lambda\rho} \, \varepsilon_{\mu\nu\lambda\rho'} 
&=& - 6 \delta^\rho_{\rho'} \,, 
\label{epseps3} \\
\varepsilon^{\mu\nu\lambda\rho} \, \varepsilon_{\mu\nu\lambda\rho} 
&=& - 24 \,.
\label{epseps4} 
\end{eqnarray}
\end{subequations}
%

\section{Note on a class of representations of Dirac
  matrices}\label{ss:gam*}
In this appendix, we want to make a comment about a class of
representations of the Dirac matrices where each matrix is either
purely real or purely imaginary.  Note that the hermiticity property
of the 16 basis matrices mentioned in \Eqn{basis16} are all determined
by their definitions and through the hermiticity property of the
$\gamma_\mu$'s given in \Eqn{gamdag}.  Thus, if any of these 16 basis
matrices has either purely real or purely imaginary elements, it would
be a symmetric or an antisymmetric matrix.  However, the number of
antisymmetric and symmetric $4\times4$ matrices must be 6 and 10
respectively.  This property, invoked already in \sec{ss:Ctrans} to
obtain the antisymmetry of the matrix $C$, can produce interesting
constraints on possible representations of the Dirac matrices of this
class.

Here is how it goes.  In this class of representation, we can
introduce the parameter $c_0$ by the definition
\begin{eqnarray}
\gamma_0^* = c_0 \gamma_0 \,.
\end{eqnarray}
In other words, if $c_0=+1$, the matrix $\gamma_0$ is real.  Since
$\gamma_0$ must be hermitian, it implies that it is symmetric in this
case.  On the other hand, if $c_0=-1$, the matrix $\gamma_0$ is
imaginary and therefore antisymmetric.  Next, we suppose that, among
the three matrices $\gamma_i$, there are $n$ matrices whose elements
are all real.  Of course $0\leq n \leq 3$.  Since the $\gamma_i$
matrices are antihermitian, this implies that $n$ of them should be
antisymmetric.  If $c_0=+1$, this means that $n$ among the three
$\sigma_{0i}$ matrices will be antisymmetric.  On the other hand, if
$c_0=-1$, we will have $3-n$ antisymmetric matrices among the
$\sigma_{0i}$'s.  Continuing the counting in this manner we obtain
that, among the 16 basis matrices given in \Eqn{basis16}, the number
of antisymmetric matrices is given by
\begin{eqnarray}
N_A = 8 - {3 \choose n} + n + \Big[ 1 + 2c_0(1-n) \Big] E \,,
\label{NA}
\end{eqnarray}
where 
\begin{eqnarray}
E = \cases {1 & if $n$ is even, \cr 
0 & if $n$ is odd.}
\end{eqnarray}
More explicitly, the result of \Eqn{NA} can be written as follows:
\begin{eqnarray}
\begin{array}{l|cccc}
& \multicolumn{4}{c}{\mbox{\shortstack{Number of
        antisymmetric\\ matrices for}}} \\ 
\mbox{$\gamma_0$ is} & n=0 & n=1 & n=2 & n=3 \\
\hline 
\mbox{real} & 10 & 6 & 6 & 10 \\
\mbox{imaginary} & 6 & 6 & 10 & 10 \\
\end{array}
\label{NAtable}
\end{eqnarray}
Of course, 10 is an inadmissible solution.  Thus, this table shows
that for real $\gamma_0$, only one or two of the $\gamma_i$'s can be
real.  On the other hand, for imaginary $\gamma_0$, the number of real
$\gamma_i$'s is either zero or one.

There is an interesting feature of the table in \Eqn{NAtable}.  It is
possible to have all $\gamma_\mu$'s to be imaginary (i.e., $n=0$ and
$c_0=-1$), but {\em not} possible to have all of them to be real (i.e.,
$n=3$ and $c_0=1$).  If all $\gamma_\mu$'s are taken to be imaginary,
the differential operator that acts on the field $\psi$ in
\Eqn{DiracEq} is real.  It shows that it is possible to have real
solutions of the Dirac equation in some representation.  Such
solutions for the field are called Majorana fields, and the
representation in which all $\gamma_\mu$'s are imaginary is called the
Majorana representation of the Dirac matrices.

However, since $n=3$ produces inadmissible solutions, all
$\gamma_\mu$'s cannot be taken to be real.  Accordingly, the matrix
multiplying the spinors in \Eqn{uveqn} cannot be real, and so the
spinors can never be purely real in any representation.


\end{document}